\documentclass[a4paper,sn-basic]{sn-jnl}%

\usepackage{graphicx}%
\usepackage{multirow}%
\usepackage{amsmath,amssymb,amsfonts}%
\usepackage{dsfont}
\usepackage{amsthm}%
\usepackage{mathrsfs}%
\usepackage[title]{appendix}%
\usepackage{xcolor}%
\usepackage{textcomp}%
\usepackage{manyfoot}%
\usepackage{booktabs}%
\usepackage{algorithm}%
\usepackage{algorithmicx}%
\usepackage{algpseudocode}%
\usepackage{listings}%
\usepackage{tikz}
\usepackage{anyfontsize} %
\usepackage{flafter}

\usepackage[normalem]{ulem} %

\usepackage[capitalize]{cleveref}
\crefformat{equation}{#2(#1)#3}
\crefformat{figure}{Figure~#2#1#3}

\theoremstyle{thmstyleone}%

\theoremstyle{thmstyletwo}%

\theoremstyle{thmstylethree}%

\graphicspath{{./figures/}}

\newif\ifshowcomments
\showcommentstrue %

\newif\ifshowcolor
\showcolortrue %

\newif\ifshowremoved

\renewcommand{\P}{\mathbb{P}}
\newcommand{\R}{\mathbb{R}}

\newcommand{\Prob}{\mathbb{P}}
\newcommand{\Y}{\mathbf{Y}}

\newcommand{\1}{\mathds{1}}

\DeclareMathOperator*{\Var}{Var}

\DeclareMathOperator*{\argmin}{arg\,min}
\newcommand{\bm}[1]{\mathbf{#1}}

\newcommand{\evacdc}{EVA (2023) Conference Data Challenge}

\begin{document}

\title[EVA (2023) Conference Data Challenge: Team genEVA]{Modeling Extreme Events: Univariate and Multivariate Data-Driven Approaches}
\subtitle{{\normalsize EVA (2023) Conference Data Challenge: Team genEVA}}

\author[1]{\fnm{Gloria} \sur{Buritic\'a}}\email{gloriapatricia.buriticaborda@agroparistech.fr} %

\author[2]{\fnm{Manuel} \sur{Hentschel}}\email{manuel.hentschel@unige.ch}

\author*[2]{\fnm{Olivier~C.} \sur{Pasche}}\email{olivier.pasche@unige.ch}

\author[3]{\fnm{Frank} \sur{R\"ottger}}\email{f.rottger@tue.nl}

\author[2]{\fnm{Zhongwei} \sur{Zhang}}\email{zhongwei.zhang@unige.ch}

\affil[1]{\orgdiv{AgroParisTech, INRAE, UMR MIA}, \orgname{Université Paris-Saclay}, \orgaddress{\city{Paris}, \country{France}}}

\affil[2]{\orgdiv{Research Center for Statistics}, \orgname{University of Geneva}, \orgaddress{\city{Geneva}, \country{Switzerland}}}

\affil[3]{\orgdiv{Department of Mathematics and Computer Science}, \orgname{Eindhoven University of Technology}, \orgaddress{\city{Eindhoven}, \country{The Netherlands}}}

\abstract{%
This article summarizes the contribution of team genEVA to the EVA (2023) Conference Data Challenge. 
The challenge comprises four individual tasks, with two focused on univariate extremes and two related to multivariate extremes. In the first univariate assignment, we estimate a conditional extremal quantile using a quantile regression approach with neural networks. 
For the second, we develop a fine-tuning procedure for improved extremal quantile estimation with a given conservative loss function. 
In the first multivariate sub-challenge, we approximate the data-generating process with a copula model. 
In the remaining task, we use clustering to separate a high-dimensional problem into approximately independent components. 
Overall, competitive results were achieved for all challenges, and our approaches for the univariate tasks yielded the most accurate quantile estimates in the competition.
}

\keywords{\evacdc{}, extreme value theory, extreme quantile regression, extremal dependence}

\maketitle

\newpage

\section{Introduction}\label{s:intro}

The \evacdc{}~\citep{EVA23Compet} comprises four individual problems in extreme value statistics, which share complications with real-world examples. 
In particular, the simulated data constructed for the challenge are compromised by issues such as missing values.
The present note provides a detailed summary of the modeling and inferential approaches of team genEVA, with particular focus on new or uncommon approaches. The team was ranked in third place overall (or in second place ex aequo before the splitting rule).

The setting for the individual sub-challenges are environmental data sets from a fictional planet.
The meteorological data, made publicly available by \cite{EVA23dat}, were simulated such that previous knowledge of climatic conditions on Earth is not useful for the analysis, and no spatial information is provided. 
There are two distinct data settings for the respective goals of predicting extreme quantiles (Tasks~C1 and C2), and estimating probabilities in relatively high-dimensions (Tasks~C3 and C4).

In C1 and C2 the task is to quantify the risk of extreme events using return levels, or extreme quantiles.
Return levels are a common risk measure in hydrology and environmental sciences where data is collected regularly, say daily, so the $T$-year return level, $q_T$, can be interpreted as the level
that the data is expected to exceed once every $T$ years.
Formally, for a univariate variable $Y$, the quantity $q_T$ can be defined by %
\begin{align}\label{eq:return:level}
q_{T} = F_Y^{-1}\left(1- \frac{1}{T n_Y }\right),
\end{align}
where $F_Y$ is the distribution of $Y$, $n_Y$ is the number of measurements collected per year, and
 $T$ is referred to as the return period.
As we keep recording new observations of the variable $Y$ on a daily basis, new records are likely to hit unprecedented levels.
In this manner, return levels are a measure of how frequently the variable $Y$ reaches high levels.
Return levels play an important role when it comes to designing risk plans; for example, daily temperature maxima or daily river discharges are of great interest to create policies aiming to protect local communities against natural hazards such as heatwaves and floods.

As the occurrence of rare events often depends on several drivers, understanding how risk varies with these is also essential for effective planning. Conditional quantile estimates are therefore an important complement to static return levels.
The conditional quantile of $Y$ given a vector of observed covariates $\bm X = \bm x$ is defined as
\begin{equation}\label{e:condquantile}%
Q_{\bm x}(\tau) =F^{-1}_{Y\mid \bm{X} = \bm{x}}(\tau), \quad \tau\in(0,1).
\end{equation}%
For extreme levels, where only a few data points are available, standard quantile regression methods~\citep{koen1978} are only of limited use for estimating conditional quantiles, as they fail to extrapolate.
This led to the development of extreme quantile regression.
Classical approaches use, for example, linear models~\citep{Wang2012, EXQAR}, generalized additive models~\citep{ExGAM,ExGAM2} or kernel methods \citep{abdelaati2011, GardesStupfler2019, Velthoenetal2019}. However, those approaches typically suffer from limitations in either flexibility, as they are not designed to capture non-linear or non-additive dependencies, or in covariate dimensionality \cite{erf}.
This motivated recent interest in more flexible and machine-learning based approaches to address more complex multivariate dependencies, for example using tree-based ensembles~\citep{gbex,erf,Kohfireboosting}, or neural networks~\citep{Pasche2022,RichardsNNfire, AlloucheNN}.
Providing reliable confidence intervals (CIs) for extreme quantiles can be a difficult task, even for static estimates~\citep{ZederPasche23}. For complex dependence models, one usually relies on bootstrap methods~\citep{DavisonBootstrap,DavisonBootstrap2} to estimate covariate dependent CIs. Although bootstrap CIs are well understood for some extreme statistics~\citep{HaanZhouExtrBoot}, this is not the case for the more flexible covariate-dependent models. One main difficulty is the sensitivity of extreme value estimates to the most extreme observations. This can, for example, lead to non-parametric bootstrap statistics exhibiting unwanted discrete behaviors on finite samples.

Tasks C3 and C4 ask for multivariate extremal probability estimates, which require extremal dependence modeling.
Here, a common assumption is multivariate regular variation \citep{resnick2008}, which allows for two related asymptotic models.
There are two classes of extremal dependence called asymptotic dependence and asymptotic independence (see \cref{s:C3} for formal definitions), where certain popular models require the assumption of the former.
The limit distribution of normalized maxima of independent and identically distributed (i.i.d.) random vectors is called a max-stable distribution \citep{Haan1984}.
Multivariate Pareto distributions are defined as the limit arising when conditioning on the exceedance of at least one variable of a high threshold, while properly renormalizing \citep{RT2006}.
Such models can only model asymptotic dependence.
Heffernan and Tawn introduced a more flexible approach that can capture both forms of extremal dependence \cite{HT2004}.
Under the assumption of asymptotic dependence, the recent introduction of extremal conditional independence \cite{eng2018} allows the definition of graphical models in extremes.
This innovation has sparked a new line of research in multivariate extremes, see e.g.~\cite{engelkeVolgushev2022,Engelke2021}.
Here, particular focus is on parsimonious graphical models for the parametric H\"usler--Reiss family of multivariate Pareto distributions, see e.g.~\cite{WZ2023, HES2022,REZ2021}.
In Tasks C3 and C4 we employ different approaches for extremal dependence modeling. For the first problem in C3, the data generating process is approximated by a Gaussian copula with Gumbel margins. This copula is designed such that its pairwise coefficients of asymptotic independence match the estimated coefficients of the data generating process.
Such an approach allows for a simple approximate calculation of the point estimate of interest.
For the second problem in C3 we rewrite the given probability as a conditional tail probability, thereby transforming the task into a univariate problem.
Task~C4 concerns a high-dimensional scenario with dimension $d=50$. We study the extremal dependence structure via a clustering approach and find five clusters of variables that are mutually asymptotically independent. Each cluster is then modeled individually, and the results are combined under the assumption of independence.

The rest of this note is structured in two major sections and finishes with a brief conclusion (\cref{s:concl}). \cref{s:quanttasks} discusses the related quantile estimation Tasks C1 and C2 and \cref{s:probtasks} the related joint-probability Tasks C3 and C4.
We give a detailed introduction to the respective settings and modeling approaches in each section. 
The open-source code and instructions to reproduce our results are available
as a github repository\footnote{See \url{https://github.com/ManuelHentschel/genEVA2023}.}.

\section{Extreme Quantile Estimation}\label{s:quanttasks}

\subsection{Goals and Data}\label{s:dataC2C1}

For the first two tasks of the \evacdc{}, the competing teams were asked to analyze a synthetic dataset composed of $n=21,000$
observations $\{(\bm{x}_i,y_i)\}_{i=1}^n$ of the random vector $(\bm{X},Y)$, where $Y$ is the response variable, taking values in $\R$, and $\bm X$ is the vector of $p=8$ covariates, taking values in a set $\mathcal{X}\subset\R^{p}$.
Observations from six of the eight covariates are missing completely at random in the training set. Further details about the variables and their meaning are given in~\citet{EVA23Compet}.
The goal of the first two competition challenges was to compute univariate extreme quantiles, both conditional and unconditional on the covariates.
For a more natural presentation, we discuss the unconditional case first (Task~C2), before introducing covariate dependence (Task~C1).

For Task~C2, the aim was to obtain an estimate $\hat{q}_T$ of the static return level $q_T$ satisfying $q_{T} = F_Y^{-1}(1-(6\cdot 10^{4})^{-1})$ that minimizes the asymmetric loss~\eqref{e:lossC2}, given below. Our proposed methodology for this task is described and assessed first in \cref{s:C2}.

In addition to the training data, we had a test dataset composed of $n_t=100$ new observations of the covariate vector $\bm X$ only. The goals for Task~C1 are to estimate, for each covariate vector $\bm{x}$ in the test set,
\begin{enumerate}
\item[a.] the {conditional quantile} of $Y$, $Q_{\bm x}(\tau)$, at the {extreme} level $\tau = 0.9999$, and
\item[b.] the corresponding central $50\%$ {confidence interval} (CI) for $Q_{\bm x}(\tau)$.
\end{enumerate}
In the data competition, the performance for Task~C1 was only assessed by how close the CI coverage of the true test response values was to 50\%, with the average CI width used only as a tie-breaker~\citep{EVA23Compet}. Even though a perfect score is achievable by producing 50 very narrow and 50 very wide intervals, we focus on a methodology providing genuine 50\% confidence intervals that are meaningful for each individual observation.
Our proposed approach for this two-part task is discussed in \cref{s:c1met}.

\subsection{Univariate Quantile Inference (Task~C2)}\label{s:C2}

The goal of C2 is to obtain reliable point estimates of high return levels $q_T$ for the response variable $Y$ which minimize the asymmetric loss function:  $\hat q_T \mapsto L(q_T, \hat q_T) $, where
\begin{equation}\label{e:lossC2}
      L(q_T, \hat{q}_T) =
          \begin{cases}
          0.9(0.99q_T - \hat{q}_T), & \text{ if } 0.99q_T > \hat{q}_T, \\
          0, & \text{ if } |q_T-\hat{q}_T| \leq 0.01 q_T, \\
          0.1(\hat{q}_T-1.01q_T), & \text{ if } 1.01q_T < \hat{q}_T,
          \end{cases}
\end{equation}
for $T \in (0,\infty)$, where $\hat q_T$ is an estimate of the true return level $q_T$; see \eqref{eq:return:level}.
The asymmetric loss function in \cref{e:lossC2} is designed to penalize underestimation more than overestimation.
It was provided by the organizers of the \evacdc{} and all teams knew in advance that Task~C2 would be scored using this loss function.
Naturally, a full risk study is the result of a balancing equation between the level of protection aimed for, and the cost associated with it.
Overestimating the level typically incurs excessive expenses, and a misallocation of resources, but underestimating the risk can have significant consequences.
In real life applications, we may be willing to accept extra costs if this means additional guarantees that the impact of catastrophes will be contained, which makes the nature of the problem asymmetric.

\begin{figure}[tb]
	\centering
	\includegraphics[width=.5\textwidth]{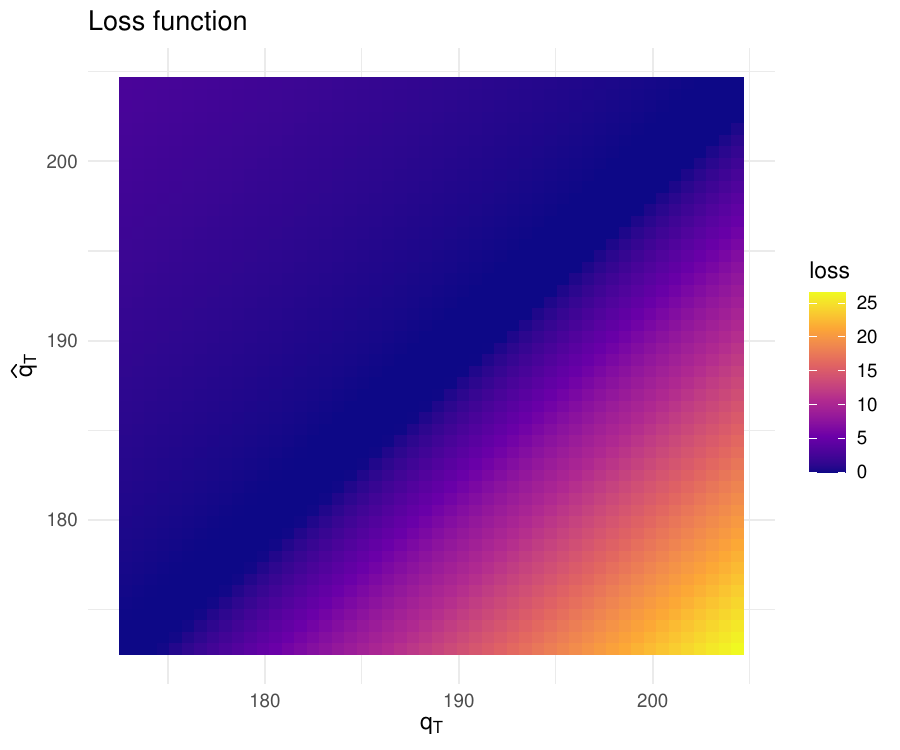}
	\caption{
		Heatmap of the function $L : \mathbb{R}^2 \to \mathbb{R}$ in \eqref{e:lossC2} evaluated on
		points $\hat q_T \in  [ \hat q_T^\ell, \hat q_T^u]$
		where $[ \hat q_T^\ell, \hat q_T^u]$ is a confidence interval for the $T$-year return level $q_T$
		of the variable $Y$.
		Notice we can't compute the exact values of the loss function derived from \eqref{e:lossC2} because the true value $q_T$ is unknown. 
	}
	\label{eq:fig:loss}
\end{figure}

Let $Y_1, \dots, Y_n$ be the data set provided for the \evacdc{}. Relying on classical statistical methods for quantile inference, it is possible to construct a sample-based
confidence interval of the $T$-year return level $[\hat q_T^l,\hat q_T^u]$ where we denote by $\hat q_T^l $ and $\hat q_T^u$ the lower and upper bounds of the confidence interval.
This interval gives a range of plausible values of the $T$-year return level that can be used as a starting point for tuning our inference.
To choose a point estimate of $q_T$ within this confidence interval that takes into account the asymmetric loss function of interest,
we propose to solve the equation
\begin{align}\label{eq:minim}
\hat q_T^{op}(\lambda) = \underbrace{\left\{
		\argmin_{q \in [ \hat q_T^l, \hat q_T^u ]   }\int_{\hat q_T^l}^{\hat q_T^u} L(s,q) \, ds
	\right\}}_{ \text{expected loss} } -
	\underbrace{
	\vphantom{ \left\{ \argmin_{ q \in [ \hat q_T^l, \hat q_T^u ]   }\int_{q_T^l}^{q_T^u} \right\} } \lambda 
	(\hat q_T^u - \hat q_T^l)
	}_{\text{penalty term} },
\end{align}
for  some $\lambda \geq 0$ to be selected, where
$L:\mathbb{R}^2 \to \mathbb{R}$ is the function in \eqref{e:lossC2}.
To illustrate this tuning strategy, we plot in \cref{eq:fig:loss} the function $L$ evaluated at a range of possible values for $q_T$ and estimates $\hat q_T$ taken from the confidence interval.
Notice we cannot actually compute the true values of the loss function $\hat q_T \mapsto L(q_T, \hat q_T )$ because the true value $q_T$ is unknown, hence we instead demonstrate it for a range of possible values for $q_T$.
We see from \cref{eq:fig:loss} that the higher values  in the confidence interval of $\hat q_T$ have a smaller expected loss compared to lower values of the confidence interval; see \eqref{eq:minim}.
The parameter $\lambda$ in \eqref{eq:minim} can be seen as a tuning parameter, when $\lambda = 0$, it means we look for a safe estimate in the upper part of the confidence interval, but if $\lambda > 0$, we opt for a lower estimate, which can be seen as a riskier choice because its expected asymmetric loss in \eqref{eq:minim} will be higher compared to the case where $\lambda = 0$.
To implement our strategy, we first need to compute the $T$-year return level confidence interval from data, and then we need to fine-tune
the parameter $\lambda$ appearing in \eqref{eq:minim}.
These two steps allow us to compute a point estimate of the $T$-year return level by solving Equation~\eqref{eq:minim}.
In the remainder of this section we discuss in detail how we have undertaken these two steps for the \evacdc{}.

To obtain a confidence interval of a high return level we rely on univariate extreme value statistics \citep[cf.][]{Coles2001} and consider the peaks-over-threshold method,
where, given a high threshold $u$,
we fit a generalized Pareto distribution to the exceeded amounts, and then compute an estimate of the $T$-year return level as
\begin{align}\label{eq:q:gpd}
\hat q_T = u + \frac{\hat \sigma}{ \hat \xi} \left\{ \left(\frac{n_u}{T \cdot n_Y \cdot n}\right)^{-\hat \xi} -1 \right \},
\end{align}
where $n_Y$ is the number of observations recorded per year, $n_u = \sum_{i=1}^n\1(Y_i > u )$  and $\hat \sigma, \hat \xi$ are the parameters of the fitted generalized Pareto distribution.
In the context of the \evacdc{}, we obtained these parameters using profile maximum likelihood methods, and then used them to compute $95\%$-confidence intervals of the $T$-year return level $q_T$ using the Delta-method \cite{Coles2001}.

We focus on estimating the $200$-year return level $q_{200}$ of the response variable $Y$ based on $70$ years of daily records, 
and recall that for the \evacdc{} we assume one year has $300$ days yielding a sample size of $n = 21,000$ observations. 
Since the sample size is reasonably large, we can fine-tune the value of $\lambda$ by performing the $k$-fold cross-validation procedure that we describe next.
We divide our sample into $k$ subsamples, each of length $\lfloor n/k \rfloor $, and compute estimates of the $200$-year return level based on $70$ years of daily records from a sample of length $\lfloor n/k \rfloor $ such that one year has now $300/k$ days. More precisely, this means that
for each subsample we compute a $T$-year return level confidence interval $[\hat q^l_{T,i}, \hat q^u_{T,i}] $  with $T = 200$, and $n_Y =  300/k$, for $i = 1, \dots, k$, with the notation in \eqref{eq:return:level}. 
In this case, our cross-validation approach mimics an objective that is of a similar difficulty as the one in challenge C2.
Next, we use the observations that do not belong to the $i$-th sample to compute the empirical $T$-year return level that we denote $q_{T,i}$, and which we use as an approximation of the true value. Considering a larger number of folds $k$ will yield a greater accuracy of this empirical estimate.
\par 
The $k$-fold validation is introduced to
correct the bias arising from Pareto-based methods.
More precisely, choosing a low threshold $u$ in \eqref{eq:q:gpd} can produce biased estimates due to misspecification of the model when we approximate the distribution of exceeded amounts above a high threshold with a generalized Pareto distribution. Instead choosing a very high threshold $u$ decreases the effective sample size, which points to the bias-variance trade-off in extreme value statistics.
On the other hand, the empirical quantiles always yield unbiased estimates of the $T$-year return level and thus can be used to help us fine-tune our point estimates.
Our approach is to concentrate on optimizing the downstream task of obtaining unbiased estimates of the $T$-year return level instead of focusing on choosing the correct threshold $u$.
This motivates the penalty term that we introduce in \eqref{eq:minim}.
We use the empirical $T$-year return level $q_{T,i}$ to compute $\lambda_i$ that we define to be the solution to the equation
\begin{align}\label{eq:lambda:def}
\hat q_T^{op}(\lambda_i) = q_{T,i},
\end{align}
with the notation in \eqref{eq:minim}, for $i =1, \dots, k$.
Finally, we let
$\lambda_{op}$ equal the median of the obtained values $\lambda_1, \dots, \lambda_k$, and we return $\hat q_{200}^{op}(\lambda_{op})$ as the point estimate of the $200$-year return value.

\subsubsection{Methodology Assessment}

To assess the performance of our fine-tuning algorithm we conduct a small numerical experiment which aims to compare the classical high return level point estimate obtained from the generalized Pareto distribution fit in \eqref{eq:q:gpd}, and
our fine-tuned estimate obtained by solving \eqref{eq:minim}.
In our experiments we consider different distributions in the domain of attraction of the generalized Pareto distribution, and for each distribution we also consider different sample sizes $n$. For a fixed sample length $n$, we estimate the quantile $q$ such that $q = F_Y^{-1}(1-{70}/{(200 \, n)})$.
To obtain a classical point estimate of this quantity we fit the generalized Pareto distribution to the amounts exceeding the $95$th empirical percentile.
To implement our approach from Equation \eqref{eq:minim}, we first compute confidence intervals based on the previous Pareto fit, and then we fine-tune the value of $\lambda$ in \eqref{eq:minim} using the $k$-fold cross-validation strategy that we explained at the end of \cref{s:C2} (see \eqref{eq:lambda:def}).
In this final step we set $k = 7$, as this was also the choice we made in the final submission for the \evacdc{}.
Finally, for the fine-tuned estimate we return the point estimate obtained by solving \eqref{eq:minim} for the fine-tuned value $\lambda_{op}$ from the previous step.

\begin{table}[tb]
\begin{tabular}{l c c c c rrrr} \hline\hline
Distribution & method &\multicolumn{3}{c}{expected asymmetric loss}\\
 & &$n=5\,000$&$n=7\,000$&$n=9\,000$
\\ [0.5ex]
\hline
&classic
	&$38.2_{\color{gray}(\pm 19.4)} \cdot 10^3$
	&$84.5_{\color{gray}(\pm 69.2)} \cdot 10^3$
	&$54.0_{\color{gray}(\pm 12.8)} \cdot 10^3$ \\[-1ex]
\raisebox{1.5ex}{Fr\'echet}
\raisebox{1.5ex}
&fine-tuned
	&${\bf 14.5}_{\color{gray}(\pm 1.6)}\cdot 10^3$
	&${\bf 15.9}_{\color{gray}(\pm 1.9)}\cdot 10^3$
	&${\bf 17.0}_{\color{gray}(\pm 1.8)}\cdot 10^3$\\[1ex]
\hline %
&classic
	&$19.8_{\color{gray}(\pm 2.7 )}\cdot 10^{-2}$
	&$16.6_{\color{gray}(\pm 2.6)}\cdot 10^{-2}$
	&$16.1_{\color{gray}(\pm 2.0)}\cdot 10^{-2}$ \\[-1ex]
\raisebox{1.5ex}{normal}
\raisebox{1.5ex}
&fine-tuned
	&${\bf 16.3}_{\color{gray}(\pm 2.6)} \cdot 10^{-2}$
	&${\bf 11.7}_{\color{gray}(\pm 1.1)}\cdot 10^{-2}$
	&${\bf 9.4}_{\color{gray}(\pm 1.2)}\cdot 10^{-2}$\\[1ex]
\hline %
&classic
	&$37.6_{\color{gray}(\pm 6.8)} \cdot 10^{-1}$
	&$28.7_{\color{gray}(\pm 4.5)}\cdot 10^{-1}$
	&$32.1_{\color{gray}(\pm 4.5)}\cdot 10^{-1}$ \\[-1ex]
\raisebox{1.5ex}{t}
\raisebox{1.5ex}
&fine-tuned
	&${\bf 24.1}_{\color{gray}(\pm 2.5)}\cdot 10^{-1}$
	&${\bf 22.9}_{\color{gray}(\pm 2.2 )}\cdot 10^{-1}$
	&${\bf 20.1}_{\color{gray}(\pm 1.9)}\cdot 10^{-1}$\\[1ex]
\hline %
&classic
	&$34.5_{\color{gray}(\pm 6.2)}\cdot 10^{-1}$
	&$26.5_{\color{gray}(\pm 4.1)}\cdot 10^{-1}$
	&$28.5_{\color{gray}(\pm 2.9)}\cdot 10^{-1}$ \\[-1ex]
\raisebox{1.5ex}{Burr}
\raisebox{1.5ex}
&fine-tuned
	&${\bf 21.8}_{\color{gray}(\pm 2.0)}\cdot 10^{-1}$
	&${\bf 18.0}_{\color{gray}(\pm 1.7)}\cdot 10^{-1}$
	&${\bf 19.1}_{\color{gray}(\pm 1.7)}\cdot 10^{-1}$\\[1ex]
\hline %
\end{tabular}
\caption{Performance of the fine-tuning algorithm for high return level estimates. Bold entries indicate a better performance.}
\centering
\label{tab:marginal}
\end{table}

\cref{tab:marginal} reports the obtained results.
We see that fine-tuning the algorithm minimizes the expected asymmetric loss in all cases, and also helps to reduce the variance compared to the classical approach based on one single point estimate.
In particular, the method outperforms the classical approach regardless of the distribution from which we aim to estimate high quantiles.
Therefore, the takeaway message from our numerical experiments is that, even though our approach was designed in the context of the \evacdc{}, it could be used to improve point estimate inference in the general setting (smaller sample sizes, different underlying distributions) where we start with a loss function to evaluate inference.
Our extreme quantile prediction achieved the best accuracy in the data competition~\citep{EVA23Compet}.

\subsection{Extreme Quantile Regression with Confidence Intervals (Task~C1)}\label{s:c1met}

\subsubsection{Feature Engineering and Missing Values}\label{s:imputation}

In a regression context such as Task~C1, feature engineering and imputation are often a crucial part of modeling.
Meaningful variable transformations can allow simple models to compute covariate dependencies much more accurately. In addition, an accurate imputation of missing covariate values avoids the discarding of partial observations or covariates, thus increasing the effective sample size used for training the regression model and potentially improving its accuracy as a consequence.

As we choose to use flexible regression models down the line and wish to make no additional assumptions on the data-generating process, we keep the feature engineering to a minimum. The only notable variable transformation is the replacement of the wind direction covariate, expressed as an angle in radians, by the corresponding latitude and longitude components, obtained with the sine and cosine functions, respectively. All covariates are also rescaled to zero mean and unit variance, which is best practice for some of the considered models.

Values are missing for seven out of the nine variables and, in total, for around $12\%$ of the $21,000$ observations. Hence, a significant gain in usable training data is achievable by imputation.
As we know the covariate observations are missing completely at random, we can perform multivariate (conditional) imputation to estimate the missing values of the covariates. 
We considered several methods: softImpute~\citep{softImpute}, MICE~\citep{MICEImpute}, MissForest~\citep{MissForest} and missMDA~\citep{missMDA}. We choose MissForest for the final imputation as it yields the lowest mean squared error over the non-missing observations. 
Details and analyses of the imputation methods are not discussed in this paper. 

The imputed data is used throughout the remaining subsections related to Task~C1. In \cref{s:eqrn}, it is used to train the extreme quantile regression model. In \cref{s:bootCI}, we propose a novel semi-parametric approach to obtain prediction intervals by refitting the regression model. Finally, \cref{s:resC1} briefly discusses the results.

\subsubsection{Extreme Quantile Regression with Neural Networks}\label{s:eqrn}

The goal of Task~C1 is to estimate the conditional quantile~\eqref{e:condquantile} of $Y$ given the test covariate realizations of $\bm{X}$, at probability level $\tau=0.9999$.
On one hand, as $\tau$ is extreme relative to the number of available observations, extreme value statistics are essential for extrapolation.
On the other hand, accurately predicting conditional quantiles typically requires regression methods.
As motivated in \cref{s:intro}, extreme quantile regression methods are, therefore, a natural choice for estimating~\eqref{e:condquantile} in this context.

We wish to avoid assumptions on the dependencies between the covariates and the response variable. Thus, more flexible, often machine-learning based, methods are the most relevant choice for accurate extreme quantile prediction. Among them, we consider several methods that have been shown to perform accurately in simulation studies and have easy-to-use package implementations: generalized additive models for peaks-over-threshold~\citep[EGAM,][]{ExGAM2}, gradient boosting for extremes~\citep[GBEX,][]{gbex}, extremal random forests~\citep[ERF,][]{erf} and the independent-observations version of extreme quantile regression neural networks~\citep[EQRN,][]{Pasche2022}.
After fine-tuning each method, we select the model performing best in terms of goodness-of-fit on set-aside validation observations; see \cref{s:resC1} for more details.

We briefly describe the selected model, EQRN, which is a two-step procedure based on peaks-over-threshold~\citep{BH1974,Pickand1975} and neural networks~\citep{DL} for flexible extrapolation.

The first step is to estimate {conditional quantiles} $\hat{Q}_{\bm{x}}(\tau_0)$ at an intermediate probability level $\tau_0 < \tau$, using classical quantile regression. We use generalized random forests~\citep[GRF,][]{GRF} as they are flexible, easy-to-fit and allow for out-of-bag prediction of $\hat{Q}_{\bm{x}}(\tau_0)$ on the training set, which conveniently helps to mitigate overfitting~\citep{Pasche2022}. Using the intermediate quantiles as a conditional threshold, we obtain $n_u\approx (1-\tau_0)n$ conditional exceedances $z_i := \{y_i - \hat Q_{\bm{x}_i}(\tau_0)\}_+$, $i = 1, \dots, n_u$. Under mild distributional assumptions, the exceedances approximately follow a generalized Pareto distribution (GPD), that is
\begin{equation*}
\Prob\{Y - \hat Q_{\bm{X}}(\tau_0) \leq y \mid Y > \hat Q_{\bm{X}}(\tau_0)\} \approx 1-\left(1+\xi(\bm{x})\dfrac{y}{\sigma(\bm{x})}\right)^{-\frac{1}{\xi(\bm{x})}}_+.
\end{equation*}

The second step is to obtain predictions of the GPD parameters $(\hat{\sigma}(\bm{x}),\hat{\xi}(\bm{x}))$ with a {neural network} fitted on the exceedances, that takes as input the covariate values $\bm{x}\in\R^p$ to output the conditional parameter estimates. The network is trained using back-propagation and gradient-descent based optimization~\citep{Adam} to minimize the negative orthogonal GPD log-likelihood
\begin{equation}\label{e:ogpdlik}%
\ell_{\rm OGPD}(z;\nu,\xi)= \left(1+\dfrac{1}{\xi}\right) \log\left\{1+\xi\dfrac{(\xi +1)z}{\nu}\right\} + \log(\nu) - \log(\xi +1)
\end{equation}%
over the training set, where $(\sigma(\bm x),\xi(\bm x))\mapsto (\nu(\bm x),\xi(\bm x)), \nu(\bm x):=\sigma(\bm x) (\xi(\bm x)+1)$ is a Fisher-orthogonal re-parametrization improving training stability and convergence \citep{Pasche2022}. The architecture of the neural network and other hyperparameters are selected based on the value of~\eqref{e:ogpdlik} on a set-aside validation set, by performing a grid-search.

Finally, for a test covariate observation $\bm{x}' \in \R^p$, the extreme conditional quantile estimate for $Y\mid \bm{X} = \bm{x}'$ is %
\begin{equation*}
\hat{Q}_{\bm{x}'}(\tau) = \hat{Q}_{\bm{x}'}(\tau_0) + \dfrac{\hat{\sigma}(\bm{x}')}{\hat{\xi}(\bm{x}')} \left[\left(\frac{1 - \tau}{1 - \tau_0}\right) ^{-\hat{\xi}(\bm{x}')} - 1\right].
\end{equation*}

This approach is implemented by its original authors in the R programming language as the \texttt{EQRN} package\footnote{The \texttt{EQRN} R package is available at \url{https://github.com/opasche/EQRN.}}, which is used in this study.

\subsubsection{Semi-Parametric Bootstrap for Central Quantile CIs}\label{s:bootCI}

We propose a semi-parametric bootstrap strategy to construct central confidence intervals for $Q_{\bm x}(\tau)$ around the predictions $\hat{Q}_{\bm{x}}(\tau)$ described in \cref{s:eqrn}. The aim of the strategy is to assess the conditional variance of the selected EQRN model predictions.

The first step is to draw $B$ bootstrap samples with replacement from the training data. When sampling an observation $(\bm{x}_i,y_i)$, $i=1,\ldots,n$, the response observation $y_i$ is kept if $y_i<\hat{Q}_{\bm{x}_i}(\tau_0)$, or replaced by $y_i'\sim {\rm GPD}(\hat{\sigma}(\bm{x}_i),\hat{\xi}(\bm{x}_i))$ generated from the conditional EQRN tail GPD distribution otherwise. This yields $B$ ``semi-parametric'' bootstrap samples. The EQRN model described in \cref{s:eqrn} is then fitted again on each of these $B$ samples separately.

Then, for each test point $\bm{x}_i$, $i=1,\ldots,n_t$, $B$ predictions $\hat{Q}_{\bm{x}_i}^{*1}(\tau),\ldots,\hat{Q}_{\bm{x}_i}^{*B}(\tau)$ are obtained from the bootstrap EQRN models. The variance of these $B$ predictions is computed, and used to construct a normal confidence interval, with the desired confidence level $\alpha$, around the original prediction $\hat{Q}_{\bm{x}_i}(\tau)$.
Although the $B$ predictions might not necessarily be normally distributed, the choice of normal intervals over alternatives such as the basic or percentile CIs was taken in accordance with the goal of Task~C1, which is to estimate central intervals.
To save computational power, the $B$ neural networks fitted on the bootstrap samples use warm start, with their weights initialized as the final weights from the original fit.

The motivation for this semi-parametric approach is to circumvent issues that can arise when using a non-parametric or parametric bootstrap in our context.
On one hand, extreme value models, such as the one used above, mainly rely on the largest tail observations. On finite samples, this can sometimes result in unwanted discrete behaviors from the non-parametric bootstrap statistics, which can lead to inaccurate intervals.
The semi-parametric approach aims at obtaining a smoother tail for the sample bootstrap distribution.
On the other hand, directly drawing parametric bootstrap samples from the fitted conditional GPD would ignore the uncertainty from fitting the intermediate quantile model $\hat{Q}_{\bm{x}}(\tau_0)$, which could greatly underestimate the variance of the whole approach. The semi-parametric approach allows refitting the whole model, including $\hat{Q}_{\bm{x}}(\tau_0)$ and the GPD parameter network, to factor in its cumulative uncertainty.

\subsubsection{Results and Discussion for Task~C1}\label{s:resC1}

For each of the considered tail models (EGAM, GBEX, ERF, EQRN), we use the same GRF quantile predictions as the threshold, as motivated in \cref{s:eqrn}. We also add $\hat{Q}_{\bm{x}}(\tau_0)$ as an additional input covariate for all tail models, as this was shown to consistently improve the accuracy of every considered GPD regression model and data scenario in~\citet{Pasche2022}. For the relevant models, the hyperparameters are selected based on the loss~\eqref{e:ogpdlik} evaluated on set-aside validation data by performing a grid-search. 
We choose to use a single validation split over cross-validation for a more efficient grid-search, thanks to its lower computational complexity. 
For EQRN a varying shape parameter did not significantly reduce the validation loss, so a constant estimate was enforced. The selected tree depths in GBEX also suggest a constant shape parameter.

The best performing of these models, in terms of the validation loss, is EQRN. GBEX achieved a close second-best score, and the other two models resulted in a significantly worse validation loss. The shape parameter estimate from GBEX is slightly positive, which seems unrealistic as the static GPD shape estimate is negative, meaning that the conditional tail of $Y\mid\bm{X}$ is estimated to be heavy although $Y$ is unconditionally light-tailed. As the EQRN shape estimate is also negative, this further supports EQRN being the best fit.
The selected EQRN network has two hidden layers with 20 and 10 neurons, respectively. It uses hyperbolic-tangent activation functions and was regularized with $L^2$ weight penalty ($\lambda = 10^{-4}$) during training.

\begin{figure}[tb]
\centering
\includegraphics[width=0.5\textwidth]{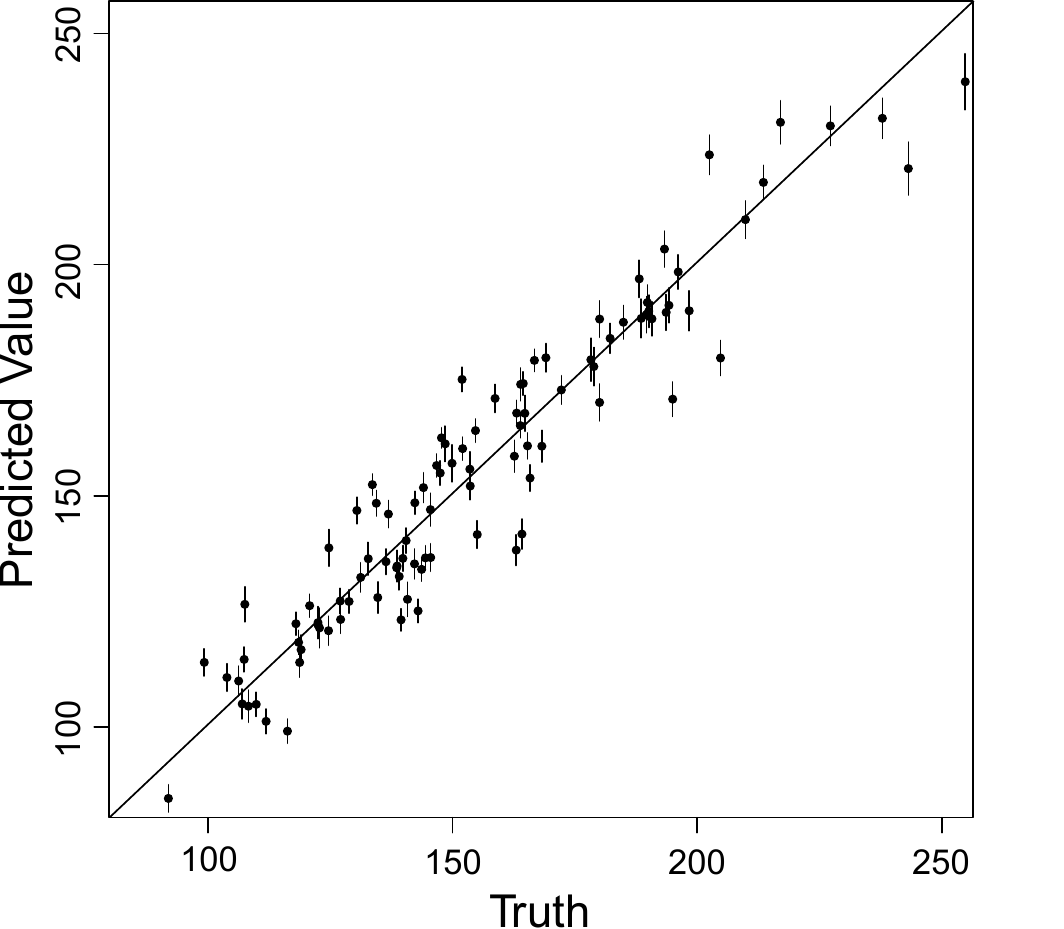}
\caption{%
    Predicted conditional quantiles $\hat{Q}_{\bm{x}}(\tau)$ on the test dataset
    compared to the truth,
    with the 50\% semi-parametric confidence intervals;
    from~\citet{EVA23Compet}.
}
\label{f:C1Cres}
\end{figure}

The EQRN conditional extreme quantile predictions achieved the best accuracy in the data competition~\citep{EVA23Compet}. \cref{f:C1Cres} compares the predictions to the truth on the test set. The model seems well calibrated and unbiased for the whole range of test quantiles.
However, the semi-parametric bootstrap procedure seems to underestimate the variance of the EQRN model, as the obtained test-set coverage of 30\% is significantly lower than the target 50\%. Although they seem to be underestimated for the whole range of values, \cref{f:C1Cres} shows that the procedure still provides intervals of varying widths with generally larger estimated variance for the largest quantiles. A possible reason for this underestimation could be the warm-start initialization of the neural networks, when re-fitted on the bootstrap samples, as they could encourage solutions in the non-convex optimization space which are similar to the original fit. However, not resorting to warm-start would significantly increase the computational cost of the procedure. Another explanation could be that the desired smoothing effect of the semi-parametric re-sampling comes at the cost of also artificially reducing the variance of the resulting bootstrap estimates. A comparison study with non-parametric bootstrap CIs could be a way to assess whether this is the case.
These results and the overall coverage performance of the other approaches in the data competition highlight that covariate-dependent confidence intervals for extreme quantiles are a challenging task, which has been given little consideration, especially for flexible tail regression models.

\section{Extremal Dependence Structure Estimation}\label{s:probtasks}

\subsection{Goals and Data}
\label{sec:c3c4goaldAndData}
In Tasks C1 and C2, interest is in the observations of $Y$ at a single site only.
In contrast, the interest in C3 and C4 lies in extremal dependence estimation, i.e., estimation of the joint probabilities of several variables (or some of them) being extreme simultaneously.
More specifically, in C3 observations of the variable $Y_i$ are available for three different locations $i = 1, 2, 3$,
as well as two of the covariates, which are meant to represent season and atmospheric conditions.
Importantly, in this challenge, the marginal distribution of $Y$ is known to be standard Gumbel everywhere over space and time.
The estimation problems in C3 thus concern solely the extremal dependence structure of the variable $Y$ at different sites.
Point estimates of the following two probabilities are required by the government of the fictional planet,
\begin{enumerate}[(i)]
    \item $p_1 := \Prob(Y_1>6, Y_2>6, Y_3>6)$;
    \item $p_2 := \Prob(Y_1>7, Y_2>7, Y_3<m)$;
\end{enumerate}
where $m=-\log(\log 2)$ is the median of the standard Gumbel distribution.

In Task~C4,
a distinct set of observations is considered,
coming from $50$ sites,
partitioned into two regions of $25$ sites each.
The quantities of interest are
the probabilities of joint threshold exceedances
with respect to different thresholds.
More specifically,
two scenarios are considered.
Below, let $s(i)$ denote the relevant threshold for site $i$,
i.e.,
$s(i) = s_1 = 5.702113$ for $i = 1, \dots 25$ and
$s(i) = s_2 = 3.198534$ for $i = 26, \dots, 50$.
\begin{itemize}
    \item Scenario i: There are two different thresholds,
        one for each region:
        \begin{align*}
            p_1 := \P(Y_i > s(i), i=1, \dots, 50)
            .
        \end{align*}
    \item Scenario ii: Only the higher threshold, $s_1$,
        is considered for all sites:
        \begin{align*}
            p_2 := \P(Y_i > s_1, i=1, \dots, 50)
            .
        \end{align*}
\end{itemize}

In the following, we describe in detail our methodologies for Tasks C3 and C4 in \cref{s:C3,s:C4}, respectively.

\subsection{Extreme Event Probability Estimation (Task~C3)}\label{s:C3}

\subsubsection{Exploratory Data Analysis}
Before selecting a parametric model for the data, we first examine the level of extremal dependence in the data by exploratory data analysis.
Here we consider the commonly used bivariate extremal dependence measures $\chi$ and $\Bar{\chi}$ \citep{Coles1999DependenceMF}.
Suppose $Y_i$, $Y_j$, $i\neq j$ have marginal distribution function $F_i$ and $F_j$, respectively, then their extremal dependence can be summarized by the coefficients
\begin{align}
    \chi_{ij} &= \lim_{u\uparrow 1} \chi_{ij}(u) = \lim_{u\uparrow 1} \Prob(F_i(Y_i)>u, F_j(Y_j)>u)/(1-u), \label{eq:extr_corr} \\
    \Bar{\chi}_{ij} &= \lim_{u\uparrow 1} \Bar{\chi}_{ij}(u) = \lim_{u\uparrow 1} 2\log(1-u)/\log\Prob(F_i(Y_i)>u, F_j(Y_j)>u) - 1.
    \nonumber
\end{align}
We call $Y_i$ and $Y_j$ asymptotically dependent (AD) if $\chi_{ij}>0$, and asymptotically independent (AI) if $\chi_{ij}=0$.
When $\chi_{ij}>0$, $\Bar{\chi}_{ij}$ is necessarily $1$, whilst $\chi_{ij}=0$ implies that $\Bar{\chi}_{ij}\in[-1,1]$.
Hence, $\chi_{ij}$ ($\Bar{\chi}_{ij}$) is particularly useful for describing extremal dependence in the presence of AD (AI) and is called the coefficient of AD (AI).

\begin{figure}[tb]
    \centering
    \includegraphics[width=0.8\textwidth]{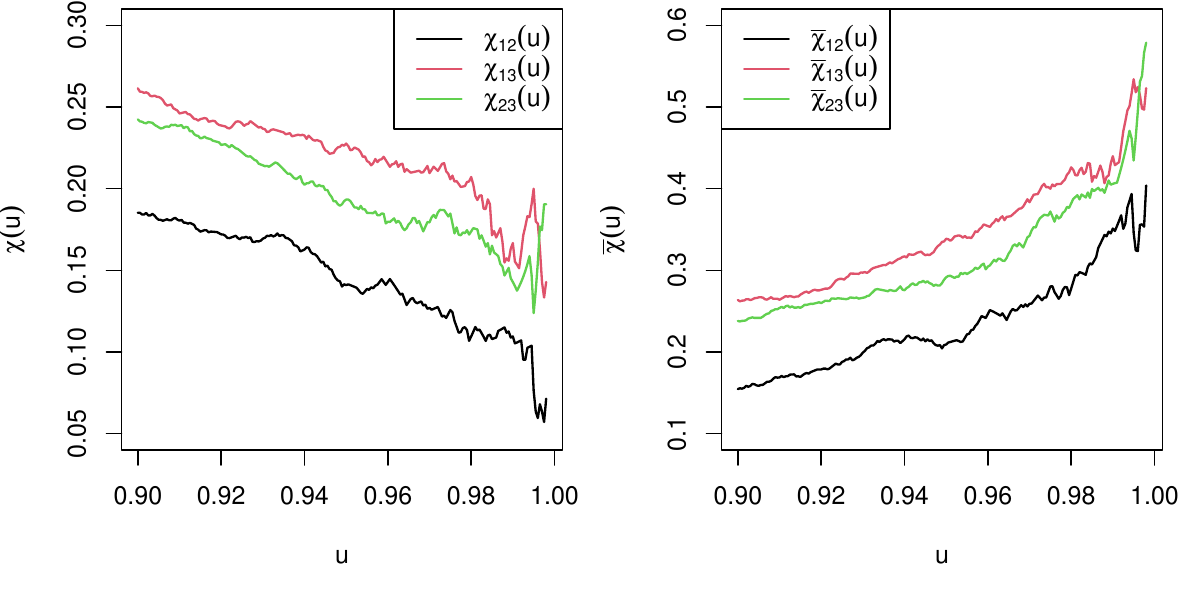}
    \caption{Empirical extremal dependence coefficients for data in C3}
    \label{fig:EDA_C3}
\end{figure}

\cref{fig:EDA_C3} depicts the empirical estimates of $\chi_{ij}(u)$ and $\Bar{\chi}_{ij}(u)$ for different pairs among $Y_1, Y_2, Y_3$.
Although it is difficult to infer whether $\lim_{u\uparrow 1} \chi_{ij}(u)=0$ for $i,j=1,2,3, i\neq j$, i.e., whether $Y_i$ and $Y_j$ are asymptotically dependent or independent, 
we do observe that the extremal dependence among all pairs seems to be weak and the level of dependence among different pairs seems to vary.
Motivated by this observation, we choose an asymptotically independent model for the estimation of the probability $p_1$; see more details in \cref{C3_p1}.
To investigate the influence of the covariates season and atmosphere on the extremal dependence structure among $Y_i$s, we consider subsets of the data with different seasons and different atmosphere levels.
The covariate season is discrete with only two possible values, S1 and S2, whilst the atmosphere is continuous.
We thus categorize the covariate atmosphere into three classes, low, middle, and high, which correspond to the lowest $30\%$, the middle $40\%$, and the largest $30\%$ of the atmosphere respectively.
The results in \cref{tab:C3_covariates} indicate that the covariates do not seem to have significant influences on the extremal dependence structure.
Hence, we choose to neglect the covariates and only use the observations of $Y$ to infer the extremal dependence structure across the three sites.

\begin{table}[tb]
\centering
\caption{Empirical extremal dependence coefficients for different seasons and atmosphere}
\begin{tabular}{|l|*{6}{c|}}\hline
    & $\chi_{12}(0.995)$ & $\chi_{13}(0.995)$ & $\chi_{23}(0.995)$ & $\Bar{\chi}_{12}(0.995)$ & $\Bar{\chi}_{13}(0.995)$ & $\Bar{\chi}_{23}(0.995)$ \\ \hline
Season S1 & $0.095$ & $0.17$ & $0.13$ & $0.39$ & $0.50$ & $0.45$ \\ \hline
Season S2 & $0.076$ & $0.21$ & $0.13$ & $0.35$ & $0.54$ & $0.45$ \\ \hline
Low atm. & $0.095$ & $0.16$ & $0.16$ & $0.38$ & $0.52$ & $0.39$ \\ \hline
Middle atm. & $0.096$ & $0.19$ & $0.096$ & $0.39$ & $0.52$ & $0.39$ \\ \hline
High atm. & $0.032$ & $0.22$ & $0.28$ & $0.21$ & $0.56$ & $0.62$ \\ \hline
\end{tabular}
\label{tab:C3_covariates}
\end{table}

\subsubsection{Estimation of Probability \texorpdfstring{$p_1$}{p1}}\label{C3_p1}
Modeling dependence for multivariate extremes is a well-known challenging problem, especially in high dimensions \citep{Engelke2021}.
Under the framework of multivariate regular variation, recent efforts to tackle this problem include the construction of sparse graphical models \citep{eng2018} and simple max-linear models \citep{Gissibl2018, Cooley2019, Kiriliouk2023}.
In the latter case, one aims to find a random vector, which follows a max-linear model, to approximate the data-generating regularly varying random vector by matching their tail pairwise dependence matrices.

Inspired by these works, we propose to find a Gaussian random vector to approximate an AI random vector by matching their matrices of pairwise coefficients of AI, $\Bar{\chi}$.
Note that we say a random vector $\Y\in\R^d$ is AI if all its marginal pairs are AI.
For $\bm{Y}\in\R^d$, we denote its matrix of AI coefficients by
$
\Bar{X}_{\bm{Y}} = (\Bar{\chi}^{\bm{Y}}_{ij})_{d\times d},
$
where $\Bar{\chi}_{ij}$ denotes the coefficient of AI of $Y_i$ and $Y_j$.
The matrix $\Bar{X}_{\bm{Y}}$ is clearly symmetric, with diagonal elements being $1$ and non-diagonal elements between $-1$ and $1$.

Our goal is to find a random vector $\bm{Z}\in\R^d$ with a Gaussian copula and standard Gumbel margins such that its matrix of AI coefficients matches that of $\bm{Y}$, i.e., $\Bar{X}_{\bm{Y}} = \Bar{X}_{\bm{Z}}$.
Assume that the matrix $\Bar{X}_{\bm{Y}}$ is positive semi-definite, then one can
first construct a random vector $\Bar{\bm{Z}}$ as $\Bar{\bm{Z}}=L\bm{W}$,
where $L$ is the lower triangular matrix associated with the Cholesky decomposition $\Bar{X}_{\bm{Y}}=LL^T$, and $\bm{W}\in\R^d$ has independent standard normal components.
Then $\bm{Z}$ can be obtained by transforming the margins of $\Bar{\bm{Z}}$ to standard Gumbel using the probability integral transform.
This construction yields that the correlation matrix ${\rm Corr}(\Bar{\bm{Z}})$ is equal to $\Bar{X}_{\bm{Y}}$, which further implies that $\Bar{X}_{\bm{Y}} = \Bar{X}_{\Bar{\bm{Z}}}$ since ${\rm Corr}(\Bar{\bm{Z}}) = \Bar{X}_{\Bar{\bm{Z}}}$ for Gaussian random vectors.
Consequently, $\Bar{X}_{\bm{Y}} = \Bar{X}_{\bm{Z}}$ since $\Bar{X}_{\bm{Z}} = \Bar{X}_{\Bar{\bm{Z}}}$.
Note that the positive semi-definiteness assumption of $\Bar{X}_{\bm{Y}}$ seems to be weak.
In the Gaussian case, it is equal to the correlation matrix and is positive semi-definite.
In general, the question of whether the matrix $\Bar{X}_{\bm{Y}}$ can only be positive semi-definite is not so straightforward to prove or disprove.
\citet{Embrechts2016} has shown the positive semi-definiteness of the matrix of AD coefficients by establishing the link between this matrix and Bernoulli-compatible matrices,
but the proof and results are not directly applicable to the matrix of AI coefficients.
Although the positive semi-definiteness of the matrix $\Bar{X}_{\bm{Y}}$ is an open question, it does not hinder the application of our approach. 
If $\Bar{X}_{\bm{Y}}$ is not positive semi-definite, one can find its nearest positive semi-definite matrix using, e.g., the algorithm of \citet{Higham2002}, and then proceed with our estimation method.

In practice, one needs to estimate the matrix $\Bar{X}_{\bm{Y}}$.
Here we use the empirical estimator of $\Bar{\chi}_{ij}^{\bm{Y}}(u)$, where $u$ is chosen to be $0.995$ since our sample size is as large as $21,000$.
Our estimate of this matrix turns out to be positive definite.
We thus use the methodology described above to obtain the random vector $\bm{Z}$ with a Gaussian copula, whose covariance matrix is specified as our estimate of $\Bar{X}_{\bm{Y}}$.
Then the probability of interest, $p_1$, can be estimated by
\[
\hat{p}_1 = \Prob(Z_1>6, Z_2>6, Z_3>6) = 2.14 \cdot 10^{-5}.
\]

\subsubsection{Estimation of Probability \texorpdfstring{$p_2$}{p2}}
To estimate the probability $p_2$, we use a different approach than the one used for estimating $p_1$ since the event of interest is not that $Y_i, i=1,2,3$ are simultaneously extreme, but rather only $Y_1, Y_2$ are extreme and $Y_3$ is smaller than its median.
Note that
\begin{align*}
    p_2 &= \Prob(Y_1>7, Y_2>7, Y_3<m) \\
        &= \Prob(Y_1>7, Y_2>7 \mid Y_3<m) \Prob(Y_3<m) \\
        &= 0.5 \Prob(Y_1>7, Y_2>7 \mid Y_3<m) \\
        &= 0.5 \Prob(\min(Y_1, Y_2) > 7 \mid Y_3<m)
\end{align*}
since $m$ is the median of the marginal distribution of $Y_3$.
Hence, estimating $p_2$ is equivalent to estimating the conditional probability $\Prob(\min(Y_1, Y_2) > 7 \mid Y_3<m)$, which boils down to estimating the tail of a random variable $\min(Y_1, Y_2) \mid Y_3<m$.

We here follow the standard approach in extreme value theory \citep{Coles2001}, and fit a generalized Pareto distribution (GPD) to high threshold exceedances of the observations of $\min(Y_1, Y_2) \mid Y_3<m$.
The threshold is chosen to be the $95\%$ quantile of the data, after examining the parameter stability plot using the R package \textit{mev}.
The estimated scale and shape parameters of the GPD are respectively $0.63$ and $0.098$.
By plugging in the estimated distribution of $\min(Y_1, Y_2) \mid Y_3<m$, we obtain the estimate of $p_2$ as $\hat{p}_2 = 3.81\cdot 10^{-5}$.

\subsubsection{Results and Discussion}
\begin{figure}[tb]
    \centering
    \includegraphics[width=0.8\textwidth]{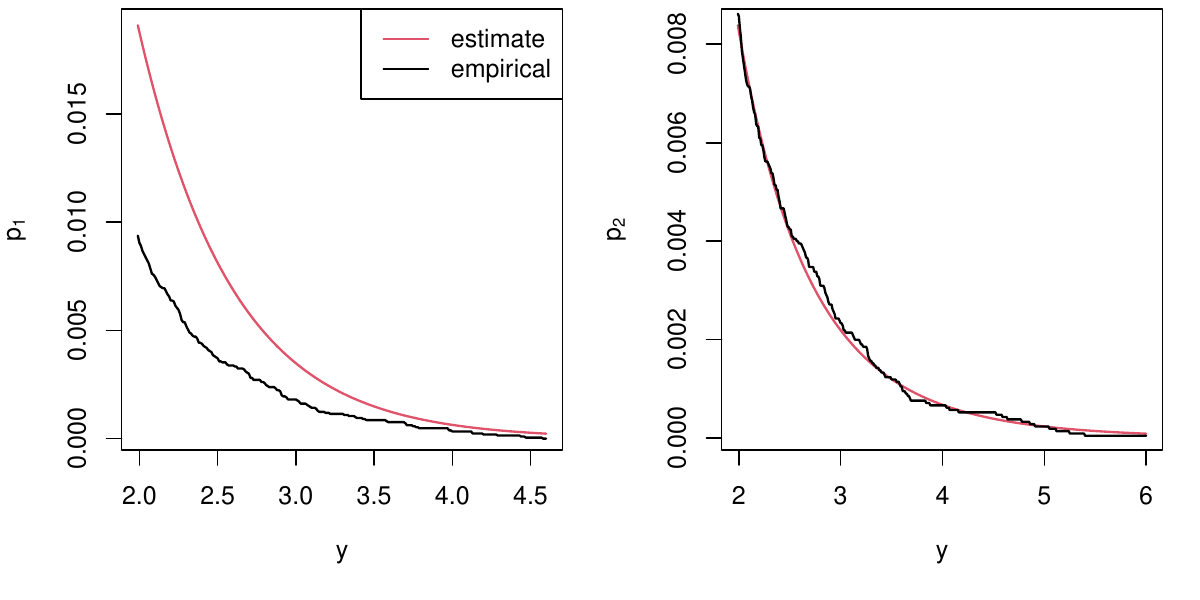}
    \caption{Empirical and our estimated probabilities of $p_1(y) = \P(Y_1 > y, Y_2 > y, Y_3 > y)$ and $p_2(y) = \P(Y_1 > y, Y_2 > y, Y_3 < m)$ for different values of $y$.}
    \label{fig:C3_diagnostic}
\end{figure}
Our estimates for probabilities $p_1$ and $p_2$ both perform well compared with other teams (see also the diagnostic plot in \cref{fig:C3_diagnostic}).
Figure~\ref{fig:C3_diagnostic} indicates that our estimate of $p_1(y)$ is slightly smaller than its empirical counterpart, while our estimate of $p_2(y)$ matches its empirical estimates much closer.
This is expected since in the first task we only used the information of AI coefficients to estimate $p_1$, but in the second task we used all the data to obtain a GPD fit of the distribution of $\min(Y_1, Y_2) \mid Y_3<m$.
Here we have shown that in the presence of weak extremal dependence, a Gaussian copula can be used as a baseline model to describe the tail of data by matching their matrix of AI coefficients.
This approach is computationally efficient and can be easily applied in high dimensions, although it is not guaranteed to correctly capture the trivariate or higher-order extremal dependence structures. 
An interesting future research question is to investigate whether the matrix of AI coefficients is positive semi-definite.
Another interesting direction is to study how to incorporate covariates information in the degree of extremal dependence.
This will allow one to construct non-stationary models for extremal dependence and capture more flexible dependence structures.

\subsection{Multivariate Threshold Exceedances (Task~C4)}\label{s:C4}

\newcommand{\plotInput}[2][0.45\textwidth]{%
    \includegraphics[width=#1]{figures/C4/#2.pdf}%
}

In Task~C4,
the probabilities of joint threshold exceedances of a relatively high-dimensional
random vector are estimated.
To structure this problem,
we employ two clustering steps,
first clustering the pairwise interactions between sites,
and then the sites themselves.
Inside each cluster of sites we model the threshold exceedance probabilities
by fitting a generalized Pareto distribution to the largest observations in the cluster.
Finally, we combine these probabilities under the assumption of
independence between different clusters,
yielding estimates of the required joint exceedance probabilities.

\subsubsection{Clustering Pairwise Dependencies}
\label{sec:clusteringPairwiseDependencies}

Since the univariate marginal distributions are known to be
standard Gumbel for each site,
it is not necessary to transform the marginal distributions of the data,
and we can turn our attention to the multivariate interactions.
We consider the matrix of pairwise
coefficients of AD $\chi_{ij}$ (as in \eqref{eq:extr_corr}) and the extremal variogram as introduced in \cite{engelkeVolgushev2022}.
The extremal variogram has proven to be very useful in the context of extremal graphical modeling,
where the entries of a random vector are associated with the node set $V$ of a graph, see \citet{eng2018,roettger2023} for further references and examples.

For a pre-asymptotic threshold $u \in (0,1)$,
the extremal variogram rooted at node $m \in V$
is a matrix with entries $i,j$ defined as
\begin{align}
    \label{eq:vario}
    \Gamma^{(m)}_{ij}(u)
    &=
    \Var(
        \log(1-F_i(Y_i))
        -
        \log(1-F_j(Y_j))
        \,\mid\,
        F_m(Y_m) > 1-u
    )
    .
\end{align}
The limiting extremal variogram can be defined by considering $u \rightarrow 1$
in the equation above.
An estimator $\hat\Gamma_{ij}^{(m)}(u)$ is obtained by replacing distribution functions with their empirical counterparts and computing the empirical variance.
Furthermore, the average over all $m \in V$ can be considered
in order to define a single estimator $\hat\Gamma_{ij}(u)$, independent of a choice of $m$,
see \cite{engelkeVolgushev2022} for details.
An empirical estimator $\hat\chi_{ij}(u)$ can be defined analogously.

\begin{figure}[tb]
    \centering
    \plotInput{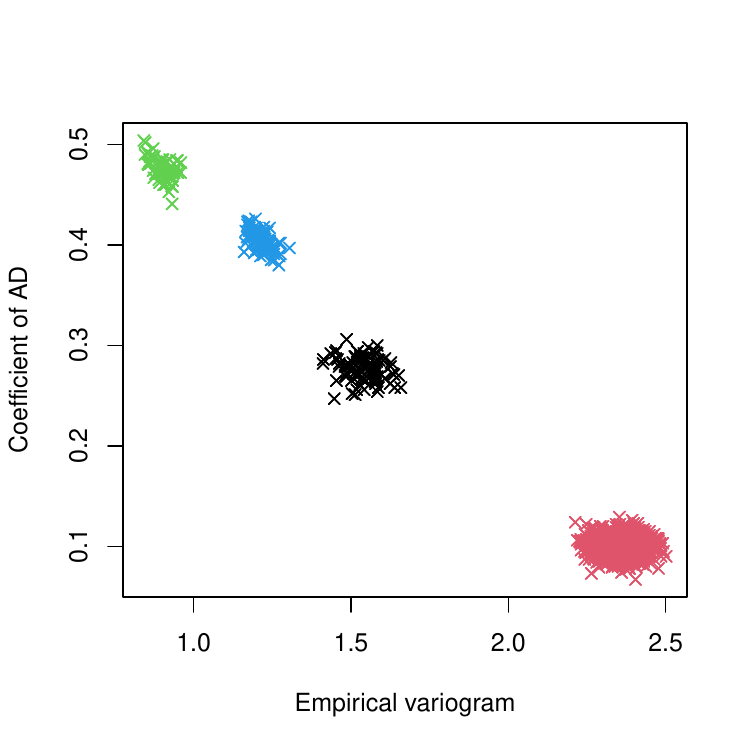}
    \plotInput{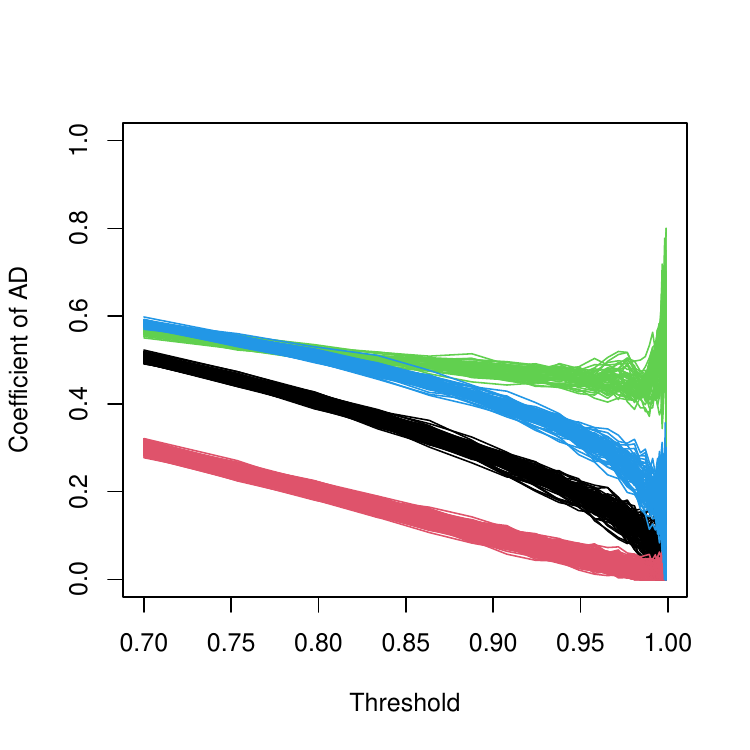}
    \caption{
        Coefficient of AD and empirical variogram
        at threshold $u=0.9$ (left)
        and over a range of different thresholds (right).
        Four distinct clusters are clearly visible,
        stable for different values of $u$,
        and are color-coded according to the results of a
        hierarchical clustering algorithm.
    }
    \label{figure:chiVsP_PreCluster}
\end{figure}

We computed the estimators $\hat\chi_{ij}(u)$ and $\hat\Gamma_{ij}(u)$
for all pairs $i,j$ and a range of threshold values $0.7 \leq u \le 1$.
The results are shown in \cref{figure:chiVsP_PreCluster}.
From the scatter plot of
$\hat\chi_{ij}(u)$ and $\hat\Gamma_{ij}(u)$
it is immediately visible that there seem to be four distinct clusters
of pairwise interactions.
The right-hand side of \cref{figure:chiVsP_PreCluster}
shows that these clusters are stable for different choices of $u$,
even though two of them overlap for $u \leq 0.85$ and
the lower two both seem to converge to zero for $u \rightarrow 1$.
One approach to use this apparent symmetry for modeling are
so-called colored graphical models,
introduced for extreme value distributions in \cite{roettger2023}.
However, as we cannot assume an underlying tree structure we would
need to assume an asymptotically dependent
H\"usler--Reiss model, which
is not feasible
as multiple clusters in \cref{figure:chiVsP_PreCluster}
appear to exhibit asymptotic independence.
Instead, we apply a more general approach, discussed below.

\subsubsection{Clustering Sites}

\begin{figure}[tb]
    \centering
    \plotInput{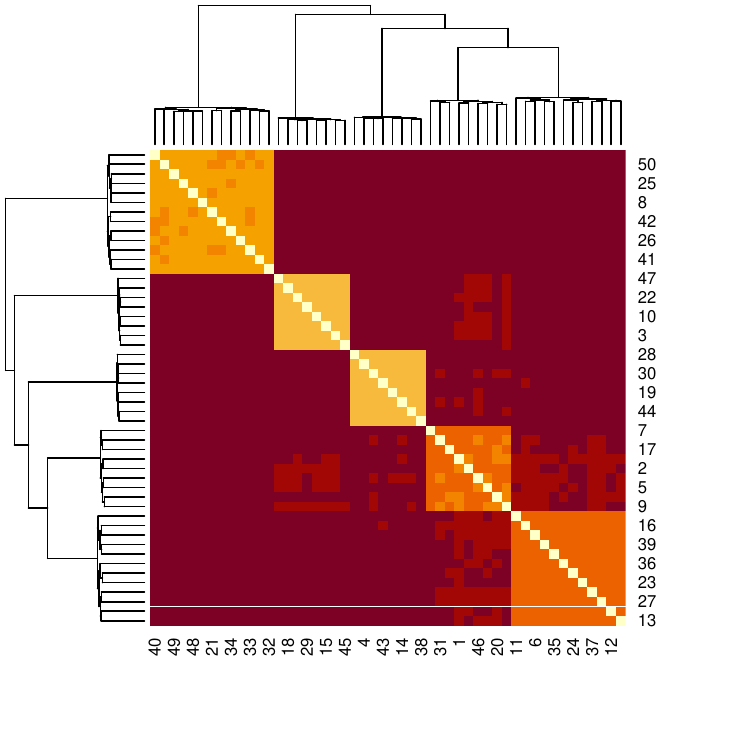}
    \plotInput{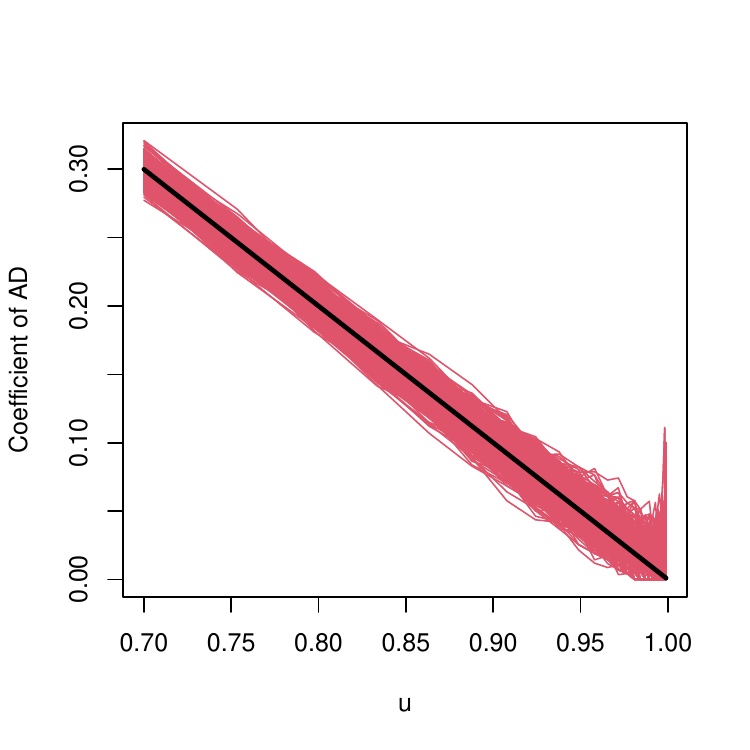}
    \caption{
        Hierarchical clustering of stations,
        using the empirical variogram $\hat\Gamma_{ij}(u)$
        at threshold $u=0.9$ as dissimilarity measure (left).
        Comparison of the
        coefficients of AD
        between disconnected vertices
        for different values of $u$,
        compared to the theoretical value for asymptotically independent variables
        (right).
    }
    \label{figure:heatmapClustering}
\end{figure}

In order to obtain a partition of the sites themselves,
rather than their pairwise interactions,
we apply a hierarchical clustering method
with the empirical extremal variogram at threshold
$u=0.9$ as dissimilarity measure,
akin to its use as distance in the minimum spanning tree
procedure described in \cite{engelkeVolgushev2022}.
The resulting dendrogram and heatmap are shown in \cref{figure:heatmapClustering},
and clearly indicate the existence of five distinct clusters of sites.
The pairwise dependencies inside each such cluster are very homogeneous,
belonging to a single cluster of interaction values for each cluster of sites.
Between sites from different clusters,
the interaction is very weak,
belonging to the large cluster of interaction values located near
zero for the coefficient of AD,
and a large value for the extremal variogram.
Note that some clusters of sites
exhibit very similar internal pairwise dependencies,
resulting in multiple such clusters
corresponding to the same cluster of pairwise interactions in \cref{figure:chiVsP_PreCluster}.
Hence, the sites are partitioned into five clusters,
and not three, as would be the case if they were in one-to-one
correspondence with the clusters of (strong) pairwise interactions.

\subsubsection{Independence Between Clusters}
In the sequel we will model the probability of joint threshold exceedances
assuming complete independence
between sites from different clusters,
yielding a simplified expression
for the probability of a joint threshold exceedance of all sites:
\begin{align*}
    \P(Y_i > s(i), i = 1, \dots, 50)
    &=
    \prod_{j = 1, \dots, 5}
    \P(Y_i > s(i), i \in C_j)
    ,
\end{align*}
where sites are indexed $1$ through $50$,
clusters are denoted $C_1, \dots, C_5$,
and $s(i)$ denotes the relevant threshold for site $i$,
depending on the considered scenario.

This assumption of asymptotic independence between clusters can be justified
since the empirical coefficients of AD $\hat\chi_{ij}$
for observations from different clusters steadily converge to zero.
Furthermore,
the classical covariance between such sites is near zero,
and the right-hand side of \cref{figure:heatmapClustering}
shows that the observed empirical coefficients of AD
(thin red lines)
are very close to the theoretical values for
independent variables (thick black line),
which can be computed to be
\begin{align*}
    \chi_{ij}(u)
    &=
    \Prob(F_i(Y_i) > u \mid F_j(Y_j) > u)
    \\ &=
    \Prob(F_i(Y_i) > u)
    \\ &=
    1 - u
    .
\end{align*}
Even though we decided to assume complete independence in the following computations,
we note that there are some indications contradicting this assumption.
For instance, in \cref{figure:heatmapClustering},
a number of empirical coefficients of AD
near $u \approx 1$
differ significantly from the theoretical value $1-u$,
and the heatmap shows that sites in the fourth cluster
(w.r.t. the order in which they are plotted)
seem to have slightly stronger dependencies with
sites from different clusters.
However,
considering how weak these indications are,
we decided to continue with the independence assumption,
leaving the modeling of inter-cluster dependencies for future research.

\subsubsection{Analysis Inside Clusters}

\begin{figure}[tb]
    \centering
    \plotInput[0.95\textwidth]{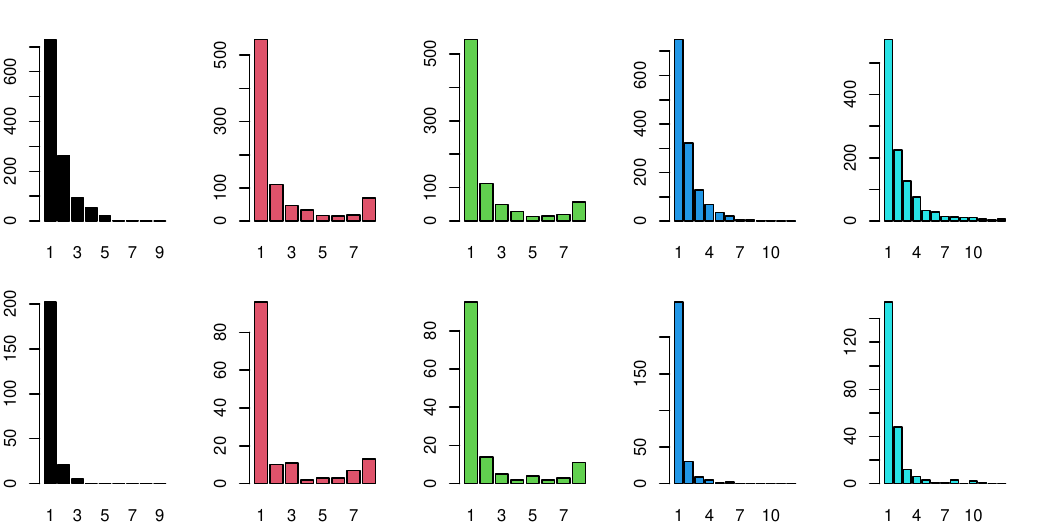}
    \caption{
        Simultaneous exceedances of the relevant threshold.
        X-axis of each plot indicates the (exact) number of observations that
        exceed the threshold at the same time.
        Y-axis indicates the number of occurrences of these exceedances.
        Top row is with respect to scenario (i),
        the bottom row is with respect to scenario (ii).
        Note the different Y-axis scales between the two rows.
    }
    \label{figure:jointExceedancesPerCluster}
\end{figure}

\cref{figure:jointExceedancesPerCluster}
shows the number of joint exceedances in each cluster
with respect to the two threshold combinations considered in the task.
For clusters 2, 3 there is a significant number of
observations in which all sites simultaneously
exceed the thresholds.
Using this,
we model the simultaneous exceedance probabilities
for these two clusters
by simply considering their empirical counterparts
\begin{align*}
    \hat\P(Y_i > s(i), i \in C_j)
    &=
    \frac{
        \#\left\{t: Y_{t,i} > s(i), i \in C_j\right\}
    }{
        T
    }
    ,
\end{align*}
for clusters $j=2,3$
and with $T$ denoting the total number of observations in the dataset.

For the remaining three clusters there are no observed simultaneous exceedances,
implying that their probabilities need to be extrapolated.
Since the empirical values of summary statistics like the coefficient of AD,
extremal variogram, and multivariate coefficient of AD \cite[cf.][]{HaanZhou2011}
differ significantly between different clusters,
we use a simple but flexible approach,
described below.

In scenario (i),
we define for each cluster index $j = 1, \dots, 5$
and threshold index $l = 1, 2$
\begin{align*}
    M_{j,l}
    =
    \min_{i \in C_j, s(i) = s_l}
    Y_i
    ,
\end{align*}
yielding for each $j$ the equality
\begin{align*}
    \P(Y_i > s(i), i \in C_j)
    &=
    \P(M_{j,1} > s_1, M_{j,2} > s_2)
    .
\end{align*}

In scenario (ii),
all stations are compared to the same threshold, $s_1$,
so the expressions above simplify to
\begin{align*}
    M_{j}
    &=
    \min_{i \in C_j}
    Y_i
    , \\
    \P(Y_i > s(i), i \in C_j)
    &=
    \P(M_{j} > s_1)
    .
\end{align*}
This is a univariate threshold exceedance,
which we model by fitting a generalized Pareto distribution (GPD)
\citep[cf.][]{Coles2001}
to the tail $T_j$ of each $M_j$.
More specifically,
we consider a high intermediate threshold $p_0$,
keep only observations exceeding the empirical quantile
$\hat Q_{M_j}(p_0)$,
and, using maximum likelihood estimation,
fit a GPD $T_{j, p_0}$ to these exceedances.
The probability of an $s_1$-exceedance can then be estimated as
\begin{align*}
    \hat\P(M_j > s_1)
    &=
    \hat\P(M_j > \hat Q_{M_j}(p_0))
    \cdot
    \hat\P(M_j > s_1 | M_j > \hat Q_{M_j}(p_0))
    \\ &=
    (1-p_0)
    \cdot
    \P(T_{j, p_0} > s_1)
    ,
\end{align*}
which can be evaluated explicitly.
In order to be more robust with respect to the choice of $p_0$,
we averaged the final probability over a range of values $p_0 \in (0.9, 1.0)$.

Coming back to scenario (i),
a bivariate exceedance probability needs to be evaluated:
\begin{align*}
    \P(M_{j,1} > s_1, M_{j,2} > s_2)
    &=
    \P(M_{j,1} > s_1)
    \P(M_{j,2} > s_2 | M_{j,1} > s_1)
    .
\end{align*}
The first factor is again a univariate exceedance probability,
which we estimate following the same procedure as above,
fitting a GPD $S_{j, p_0}$ at intermediate threshold $p_0$.
For the second factor,
we have
\begin{align*}
    \label{eq:C4bivariate}
    1
    \geq
    \P(M_{j,2} > s_2 | M_{j,1} > s_1)
    \geq
    \P(M_{j,2} > s_2 | M_{j,1} > s_2)
    =
    \chi^{(j)}(Q_{M_{j,1}}(s_2))
    ,
\end{align*}
where $\chi^{(j)}(u)$ denotes the coefficient of AD
between the two minima $M_{j,1}$ and $M_{j,2}$
at threshold $u$.
The second inequality can be justified,
since the event $\left\{M_{j,1} > s_1\right\}$
is a ``more extreme'' subset of $\left\{M_{j,1} > s_2\right\}$,
and above we have observed positive extremal dependencies between
all sites in each cluster.
Since the estimated coefficients of AD were rather large,
we estimate the probability
$\P(M_{j,2} > s_2 | M_{j,1} > s_1)$
by an estimate of its lower bound
$\chi^{(j)}(Q_{M_{j,1}}(s_2))$.

Combining the above expressions,
we get the final probability estimate
\begin{align*}
    \hat\P(M_{j,1} > s_1, M_{j,2} > s_2)
    &=
    (1-p_0)
    \cdot
    \P(S_{j,p_0} > s_1)
    \cdot
    \hat \chi^{(j)}
    ,
\end{align*}
which we again average over different thresholds $p_0$.

\subsubsection{Final Estimates and Conclusion}
\label{subsubsec:C4conclustion}

Combining the steps above,
the final estimates for the probabilities of interest are
\begin{align*}
    \hat p_1
    &=
    \prod_{j=2,3}
    \frac{
        \#\left\{t: Y_{t,i} > s(i), i \in C_j\right\}
    }{
        T
    }
    \prod_{j=1,4,5}
    (1-p_0)
    \cdot
    \P(S_{j,p_0} > s_1)
    \cdot
    \hat \chi^{(j)}
    ,
    \\
    \hat p_2
    &=
    \prod_{j=2,3}
    \frac{
        \#\left\{t: Y_{t,i} > s_1, i \in C_j\right\}
    }{
        T
    }
    \prod_{j=1,4,5}
    (1-p_0)
    \cdot
    \P(T_{j, p_0} > s_1)
    .
\end{align*}

These estimates are reasonably close to the true probabilities,
underestimating them in both scenarios.
The resulting log-scores,
which were considered for the ranking,
are very close to those of the other teams in the top five for this challenge.

Comparing to the underlying truth,
described in \cite{EVA23Compet},
it is evident that we correctly identified
the clusters used in the true data-generating process.
While the multivariate copulas inside the clusters are
widely-known parametric ones,
their parameters are random over time,
making the overall distributions
fairly non-standard.
Hence, our use of a general tail-modeling approach was justified.
However, our assumption of independence between clusters
is slightly too strong,
as there is some dependence between the copula parameters
sampled for different clusters.

We have demonstrated methods for identifying sparsity
in the structure of pairwise interactions,
and partitioning of sites accordingly.
We applied a very general, but flexible framework
to reduce the high-dimensional problem of joint threshold exceedance probabilities
to a problem of univariate and bivariate exceedances,
allowing the application of established peaks-over-threshold methods.
However,
especially in the case of mixed thresholds (scenario i),
our estimates were somewhat rough and in future research,
more complex
models could likely be employed to obtain better estimates.

\section{Conclusion}\label{s:concl}
In this section we provide a brief conclusion for each challenge and discuss potential directions for future research.

In Task~C1, we compared several models for extreme quantile regression. Our final approach, based on EQRN~\citep{Pasche2022}, achieved the best accuracy in the competition, compared to the predictions from other teams.
We also proposed a semi-parametric bootstrap approach for computing 50\% confidence intervals centered around the EQRN predictions.
The aim was to obtain smoother samples, as the non-parametric bootstrap sometimes suffers from unwanted discreteness when used for tail statistics, and to estimate the total variability of the predictive model, which would be partially ignored by a parametric bootstrap.
However, the approach seems to underestimate the variance of the model for the competition data.
Whether the underestimation is due to the use of warm-start, to the centered confidence interval or to the semi-parametric sampling remains an open question.
Further comparative studies considering random gradient-descent initialization, asymmetric percentile or basic bootstrap intervals, and non- or semi-parametric re-sampling strategies could, therefore, be valuable.

For Task~C2, we implemented a fine-tuning procedure to enhance point estimates of extreme quantiles with respect to the conservative loss function in \eqref{e:lossC2}. This loss was proposed in the context of the competition to evaluate univariate extreme quantile estimation.
Our approach yielded the best result in the competition compared to other teams, and our numerical experiment confirmed that our fine-tuning method is promising to minimize the given loss function. 

In Task~C3, we approximated the data-generating process in the first task with a Gaussian copula that matches the estimated pairwise coefficients of asymptotic independence. This resulted in a competitive estimate. It remains an open question whether more complex approximations could provide more precise estimates while maintaining the computational advantages of Gaussian copulas.

Finally, in Task~C4 we correctly identified the cluster structure of the underlying data-generating process.
Under the simplifying assumption of independence between the clusters we were able to model each cluster separately, which resulted in slightly too low, though competitive, estimates.
In retrospect, the application of more complex models that can incorporate the
identified structure
while allowing for separate modeling of the individual clusters could improve the quality of our estimates.

\backmatter

\section*{Declarations}

\bmhead{Acknowledgments}
We thank Christian Rohrbeck, Emma Simpson and Jonathan Tawn for their efforts in organizing the \evacdc{}.
We also thank the two anonymous referees for their constructive comments that have significantly improved the quality of our manuscript.
Finally, we thank Sebastian Engelke for the fruitful discussions during the data challenge.

\bmhead{Funding}
Authors were supported by Swiss National Science Foundation Eccellenza Grant 186858.

\bmhead{Availability of supporting data}
The simulated data for the \evacdc{} has been made publicly available by \cite{EVA23dat}, and is described in detail in~\cite{EVA23Compet}.

\begin{footnotesize}
\bibliography{bib_EVA_DataComp2023}%
\end{footnotesize}

\end{document}